# Deep learning-based holographic polarization microscopy

Tairan Liu[1,2,3,†], Kevin de Haan[1,2,3,†], Bijie Bai[1,2,3,†], Yair Rivenson[1,2,3], Yi Luo[1,2,3], Hongda Wang[1,2,3], David Karalli[1], Hongxiang Fu[4], Yibo Zhang[1,2,3], John FitzGerald[5], and Aydogan Ozcan[1,2,3,6,*]

[1]Electrical and Computer Engineering Department, University of California, Los Angeles, CA 90095, USA
[2]Department of Bioengineering, University of California, Los Angeles, CA 90095, USA
[3]California NanoSystems Institute, University of California, Los Angeles, CA 90095, USA
[4]Computational and systems biology department, University of California, Los Angeles, CA 90095, USA
[5]Division of Rheumatology, Department of Internal Medicine, David Geffen School of Medicine, University of California, Los Angeles, CA 90095, USA
[6]Department of Surgery, David Geffen School of Medicine, University of California, Los Angeles, CA 90095, USA
[†]Equally contributing authors
*Corresponding author: Aydogan Ozcan ozcan@ucla.edu

**Abstract**

Polarized light microscopy provides high contrast to birefringent specimen and is widely used as a diagnostic tool in pathology. However, polarization microscopy systems typically operate by analyzing images collected from two or more light paths in different states of polarization, which lead to relatively complex optical designs, high system costs or experienced technicians being required. Here, we present a deep learning-based holographic polarization microscope that is capable of obtaining quantitative birefringence retardance and orientation information of specimen from a phase recovered hologram, while only requiring the addition of one polarizer/analyzer pair to an existing holographic imaging system. Using a deep neural network, the reconstructed holographic images from a single state of polarization can be transformed into images equivalent to those captured using a single-shot computational polarized light microscope (SCPLM). Our analysis shows that a trained deep neural network can extract the birefringence information using both the sample specific morphological features as well as the holographic amplitude and phase distribution. To demonstrate the efficacy of this method, we tested it by imaging various birefringent samples including e.g., monosodium urate (MSU) and triamcinolone acetonide (TCA) crystals. Our method achieves similar results to SCPLM both qualitatively and quantitatively, and due to its simpler optical design and significantly larger field-of-view, this method has the potential to expand the access to polarization microscopy and its use for medical diagnosis in resource limited settings.

**Introduction**

Polarized light microscopy is widely used as a diagnostic tool in pathology, as it introduces distinctive contrast to birefringent specimen[1]. A number of diseases, such as squamous cell carcinoma[2], primary cutaneous amyloidosis[3], cerebral amyloid angiopathy[4], and senile cardiovascular amyloid[5] can be diagnosed using various polarization imaging techniques. Since 1961, compensated polarized light microscopy (CPLM) has been the gold standard imaging technique to identify monosodium urate (MSU)[6] crystals in synovial fluid samples[7], and is used to diagnose gout and pseudogout[8]. CPLM operates by allowing linearly polarized white light illumination to pass through a full-waveplate designed for green light (commonly between 530 nm to 560 nm), which in combination with a linear polarizer/analyzer, generates a magenta background. The presence of a birefringent specimen within the light path changes the polarization state of the green light, which shifts the spectrum after the analyzer and results in the final image becoming yellow or blue.

While CPLM images are treated as the gold standard for MSU crystal detection, the effort is labor intensive as microscopes have limited fields-of-view (FOV) and therefore, mechanical scanning is required to inspect the whole sample area. In addition to having a limited FOV, the CPLM technique suffers from limitations such as requiring manual alignment of the polarizer in relation to the analyzer, limited focal depth with higher magnification objectives and limited sensitivity when being used to detect small crystals or crystals with weak birefringence. As a result, CPLM analysis is sensitive to both the concentration of the crystals in synovial fluid[9] and experience of the diagnostician/technician[10]. Finally, clinical CPLM reporting is limited to *qualitative* results (e.g., presence or absence of crystals on the slide and whether crystals are intra- or extra-cellular).

There are a number of alternative polarization microscopy methods which have been developed to produce quantitative images of birefringent specimen. These methods all use the same principles of operation: they collect images from two or more light paths with either the polarizer or the analyzer oriented differently in order to infer the Stokes/Jones parameters that define the birefringent specimen[11–15]. Polarization holographic microscopy, (PHM)[13] one of these quantitative methods, takes advantage of the amplitude and phase information of the reconstructed interferogram, and measures spatially resolved Jones matrix components of anisotropic samples using two pairs of polarizers and analyzers. However, this system requires the use of relatively sophisticated and costly optical components to maintain a linear mapping between the measurements and the inferred Jones parameters.

Recently, a technique known as single-shot computational polarized light microscopy (SCPLM)[15] that uses a pixel-wise polarized image sensor with four polarization directions has been demonstrated to simplify the optical system required to image birefringent samples. Using this method, the retardance and orientation of the sample are explicitly solved, providing quantitative contrast for birefringent specimen. While these methods are quite effective, both PHM and SCPLM are lens-based imaging systems. Therefore, they suffer from the small field of view of objective lenses and a relatively high system cost. One method which can get past these limitations is wide-field lensfree differential holographic polarized microscopy[14,16]. By taking advantage of the simple optical design and unit magnification of lensfree on-chip holographic systems, this method can achieve a FOV of >20-30 mm$^2$ [17–19]. Furthermore, the technique is cost-effective, compact and suitable for resource-limited settings[20,21]. However, in order for this lensfree holographic imaging method to be used for imaging of birefringent objects, two sets of raw holograms must be taken with illuminations in two different polarization states, which requires precise image alignment, especially for the detection of small birefringent objects within the sample. Furthermore, this method does not resolve the retardance or orientation of the sample, leading to lower contrast compared to the SCPLM method.

In parallel to these advances in computational polarization microscopy, deep learning has emerged as a highly effective technique for solving inverse problems in microscopy[22,23]. It has been applied to traditional inverse problems such as holographic image reconstruction[24–26], reconstruction of color images[27], super-resolution[28], as well as to perform cross-modality image transformations such as virtual labeling of histological tissues[29], live cells[30], and to give brightfield image contrast to holographic images[31].

Here, we build upon these advances and present a novel deep learning-based holographic polarization microscope (DL-HPM) which can provide the retardation and orientation of birefringent specimen using a single phase-retrieved hologram that encodes one state of polarization (Figure 1). This system requires only minor changes to the existing imaging hardware, i.e., the addition of a polarizer / analyzer set to a standard lensfree holographic microscope. Our framework uses SCPLM images as the ground truth to train a deep neural network (Figure 1b), which uses this image data to learn how to transform the information encoded within a reconstructed hologram into an image that directly reveals the specimen's birefringence retardance and orientation. In addition to achieving a comparable image quality to the SCPLM images (Figure 1c), this deep learning-enabled lensfree microscopy method has a FOV of >20 mm$^2$ using a cost-effective optical design. The performance of DL-HPM is demonstrated using MSU and triamcinolone acetonide (TCA) crystal samples as well as a corn starch sample, matching the performance to SCPLM, both qualitatively and quantitatively.

We believe that the presented deep-learning based polarization microscopy approach could be widely used as a diagnostic tool in pathology and other fields that need to rapidly process and reveal the unique signatures of various birefringent crystals within complex specimen such as synovial fluid samples.

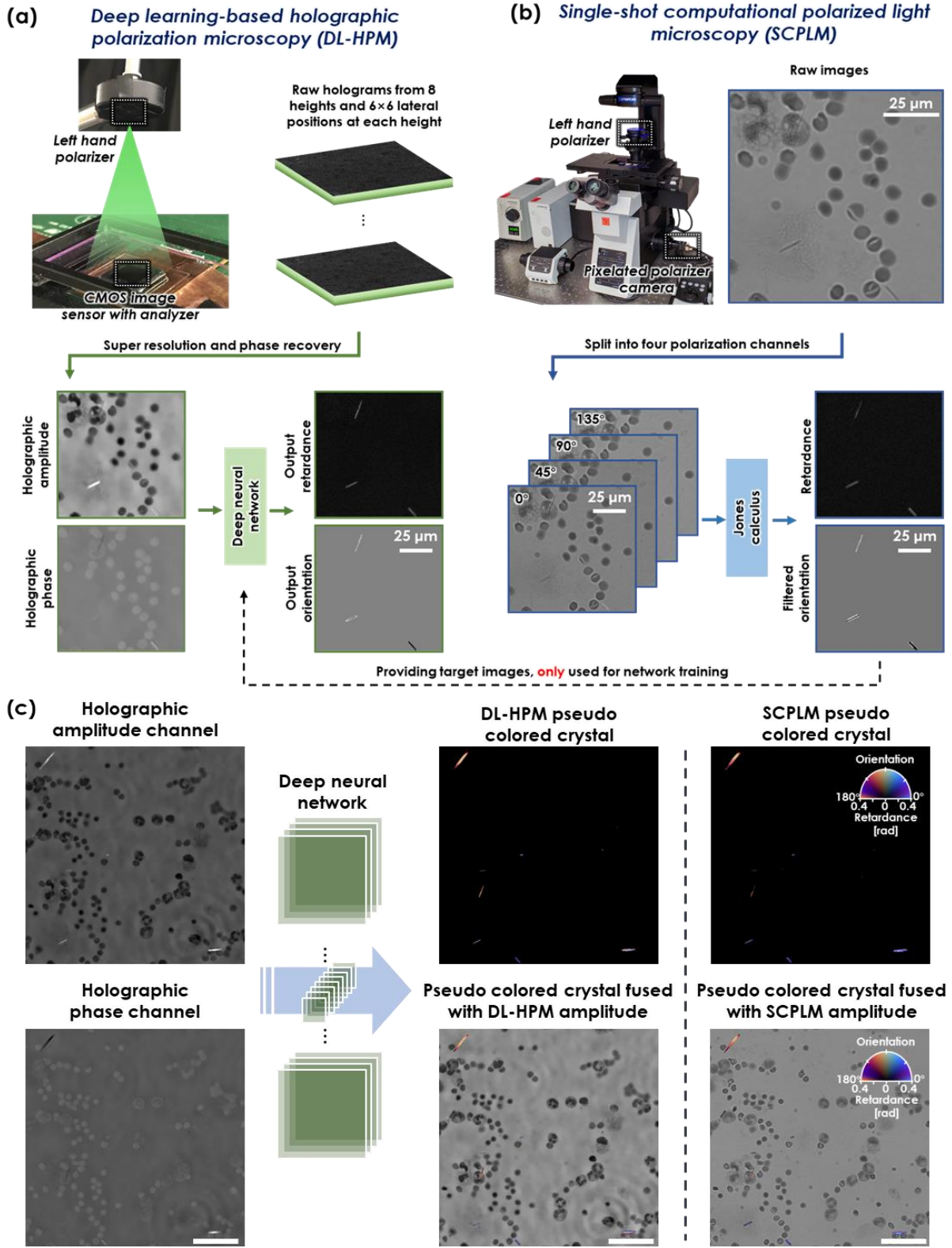

**Figure 1**. **(a) Schematic for deep learning-based holographic polarization microscopy (DL-HPM).** Raw holograms are collected using a lensfree holographic microscope with a customized polarizer and analyzer. A trained neural network is used to transform the reconstructed holographic amplitude and phase images into the birefringence retardance and orientation images. **(b) Schematic for single-shot computational polarized light microscopy (SCPLM).** Images are collected with a four-channel pixelated polarized camera under circularly polarized illumination. Birefringent retardance and orientation channels are computed using Jones calculus, and the amplitude image is obtained by averaging the four polarization channels. SCPLM is used as the ground truth information channel, providing the network training target for DL-HPM. (c) **Blind testing of DL-HPM**. A new clinical sample (containing MSU crystals) collected from a de-identified patient is tested using DL-HPM. Birefringent samples are given a pseudo color using the same convention according to the compensated polarized light microscopy. Similar image quality was achieved compared to SCPLM images. Scale bar: 50 μm.

**Results**

We trained a deep neural network (see the Methods section) using 6 clinical samples containing MSU crystals, collected from 6 de-identified patients, to perform an image transformation, from an input holographic image (amplitude and phase) to the birefringence retardance and orientation images at the output of the network. The slides were all reviewed using CPLM (Olympus BX-51) by our clinical expert (JF) for the presence of MSU crystals. This analysis found that the majority of the birefringent crystals within these samples are needle shaped MSU crystals. Once trained, the neural network was blindly tested with 2 additional MSU slides from 2 new patients; Figure 2(a) shows the blind output of the DL-HPM method in comparison to the SCPLM method. The birefringent crystals within the FOV are colored using a calibrated colormap according to the CPLM convention, where the background is left in grayscale to enhance the contrast. Figure 2(b) further shows two representative zoomed-in regions for both single MSU crystals (within a blood cell) and a crystal cluster. These images demonstrate that our deep learning framework is capable of accurately locating the birefringent objects and giving them a high color contrast with respect to the non-birefringent cell background within the FOV.

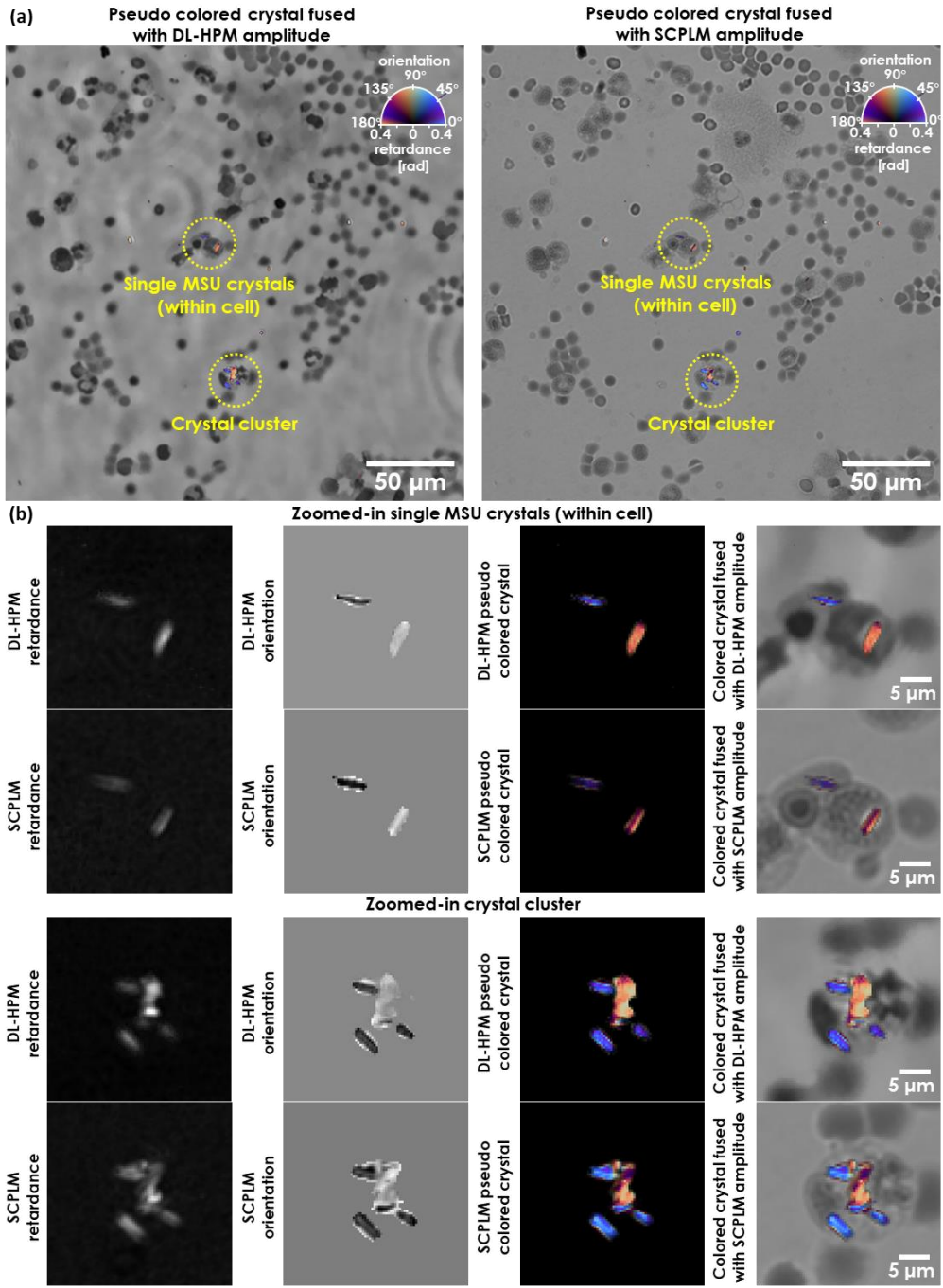

**Figure 2**. **Imaging performance of DL-HPM.** (a) Images generated using DL-HPM compared against co-registered images captured using SCPLM for a blindly tested MSU sample. The birefringent MSU crystals are colored according to CPLM convention after obtaining the retardance and orientation channels using each method. The pseudo-colored retardance and orientation information is also fused with the amplitude channel to show the high contrast against the non-birefringent cell background achieved by both methods. (b) Two different zoomed-in regions cropped from (a).

To quantify the performance of our method, 3432 different birefringent objects were detected and analyzed. These objects were composed of individual MSU crystals (most common), MSU crystal clusters, or protein clusters (rare). We classified the birefringent objects according to their length and analyzed each of these categories separately. The classification was performed by first setting a 0.2 Rad threshold on the retardance channel in SCPLM target to convert it into a binary mask. Then, using these masks, connected component analysis was performed to classify each object into different length categories. For each detected birefringent object in the SCPLM, the same coordinates were also used to locate the corresponding birefringent object in the co-registered DL-HPM. The absolute retardance/orientation error was computed pixel-wise and averaged by the number of pixels for each detected birefringent object at its local region within a 5-pixel radius of any edge of the object.

The results of this quantitative analysis are reported in Figure 3(a) with sample FOVs provided for visual comparison in Figure 3(b). The minimum length of the crystals included in this analysis was selected to be 2 μm, representing an object with at least 5 pixels in length. Objects smaller than this threshold had insufficient resolution to assign a crystal type accurately. In total, 6 length categories were selected: 2-4 μm (1077 objects), 4-6 μm (582 objects), 6-8 μm (454 objects), 8-10 μm (466 objects), 10-20 μm (795 objects), and 20-50 μm (58 objects). The error was first computed and averaged pixel-wise for each detected birefringent object, and then averaged object-wise to obtain the statistics reported in Figure 3(a). The overall object-wise averaged absolute error between the SCPLM results and the DL-HPM results was 0.047 Rad in the retardance channel and 0.135 Rad in the orientation channel, where the retardance and orientation channels range from 0 to $\pi/2$ and 0 to $\pi$, respectively. These results reveal that DL-HPM can quantitatively transform the holographic amplitude/phase information that is acquired at a single polarization state into birefringence retardance and orientation image channels, closely matching the results of SCPLM.

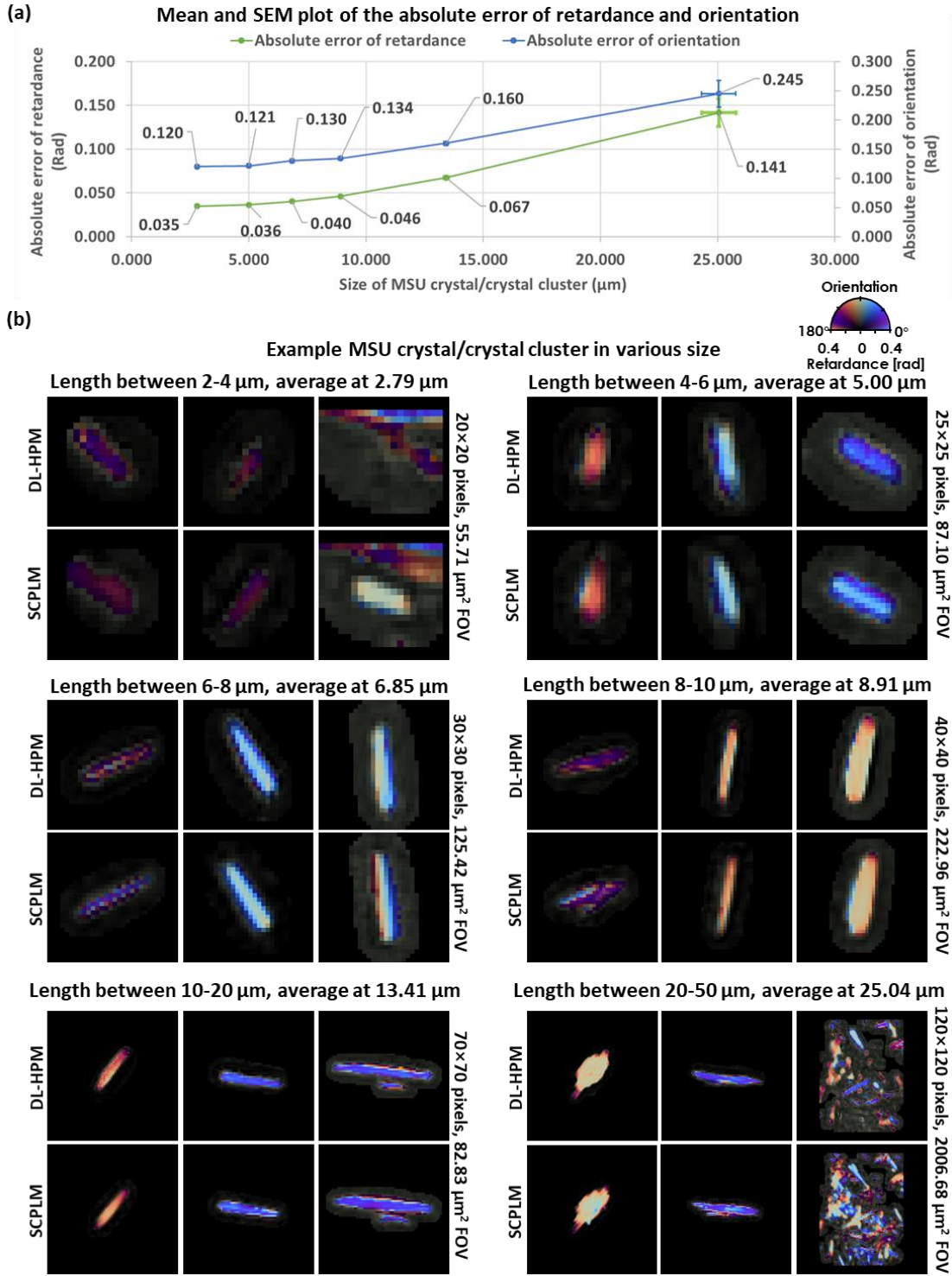

**Figure 3**. **Quantitative performance of DL-HPM, compared against SCPLM results.** (a) Mean and standard error of the mean (SEM) plots of the absolute error. DL-HPM achieves an overall object-wise averaged absolute error of 0.047 Rad in the retardance channel and 0.135 Rad in the orientation channel. (b) Sample birefringent objects. For each length category, the left image is the smallest object and the right is the largest, while the middle one has the median size.

To further investigate the image transformation performed by the trained deep network, next we blindly tested it on *two new types of birefringent samples* that were never seen by the network during its training; for this purpose, we imaged TCA crystals, and corn starch samples (Figure 4). The results revealed that DL-HPM can correctly identify birefringence in most of the TCA crystals and corn starch particles within the sample FOV. This is an indication that the presented DL-HPM method is learning a combination of the desired physical image transformation and a semantic segmentation which is related to morphological information of the samples. As expected, once the same network is further trained using transfer learning with these new types of samples, its blind inference performance can be improved, which is illustrated in Figure 5.

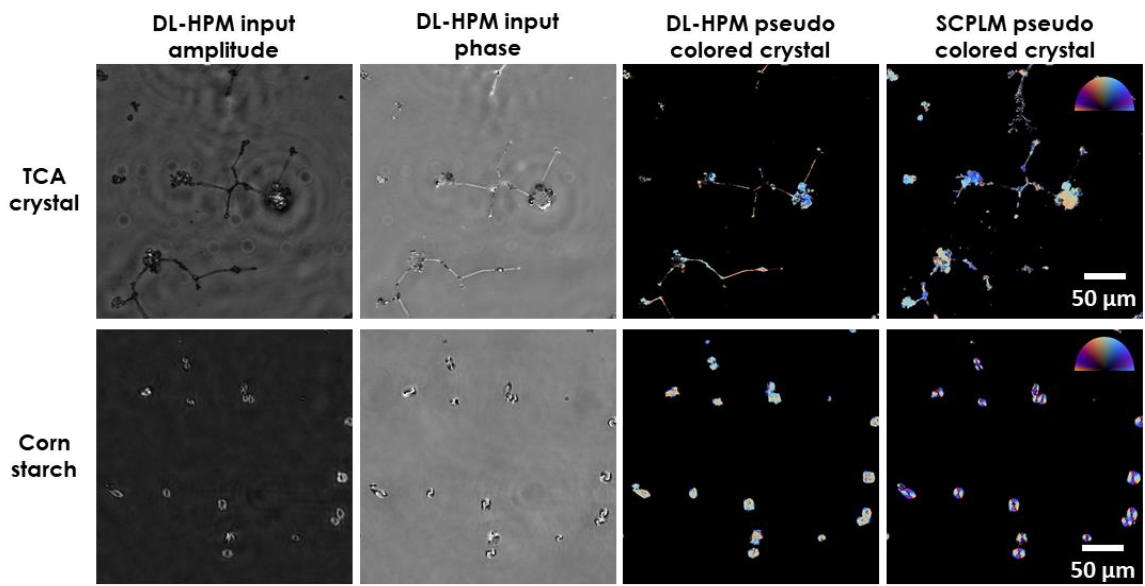

**Figure 4**. **Imaging performance of DL-HPM on new types of birefringent samples**. Visualization of birefringent TCA crystals and corn starch samples imaged using DL-HPM method, where the associated deep network is trained with only birefringent MSU crystals. Color bar: from left to top to right, represents π, π/2, 0 Rad in the orientation channel. Retardance is represented by the distance from the center of the color bar, ranging from 0 to 0.4.

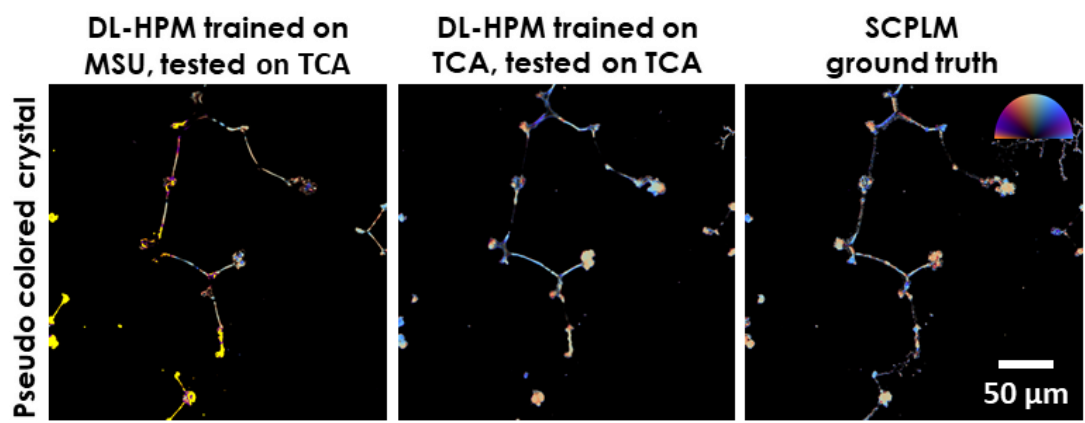

**Figure 5. Comparison of two different deep neural networks for imaging TCA samples using DL-HPM**. The first network is trained with only MSU samples, and the second network is trained through transfer learning from the original MSU network with a training dataset containing TCA samples. Color bar: from left to top to right, represents π, π/2, 0 Rad in the orientation channel. Retardance is represented by the distance from the center of the color bar, ranging from 0 to 0.4.

**Discussion**

Our results have qualitatively and quantitatively demonstrated the effectiveness of the presented framework using multiple types of samples, also illustrating the capability of the networks to generalize from one sample type to another. However, the black box nature of deep neural networks often makes it challenging to determine how the transformation is actually performed. In this section, we perform an ablation study aiming to partially reveal the physical interpretation of the deep neural network, and demonstrate that the network is learning to perform crystal segmentation based on both the morphological information and the physical relationship between the holographic amplitude/phase information and the birefringent retardance/orientation channels. For this analysis, we trained two additional networks using the MSU image dataset: 1. using only the holographic *amplitude channel* as the input to the neural network to blindly perform the retardance/orientation inference; and 2. using only the holographic *phase channel* as the input to the neural network to blindly perform the retardance/orientation inference. Examples of the blind inference performance of these trained networks are shown in Figure. 6.

In general, using only the amplitude or only the phase channel, as opposed to using both channels together, significantly degrades the inference performance of the network. The amplitude only network tends to accurately predict the crystals but generates images with significant error in the orientation channel, whereas the phase only network tends to predict the locations of the crystals less accurately. One possible explanation for this observation is that the amplitude only network is learning morphological information to locate the crystals, as the holographic optical system was designed to introduce intensity contrast for birefringent samples[14], and the additional phase information is required to fully reconstruct the orientation channel. Hence, both the amplitude and phase information channels are essential to accurately infer the retardance and orientation of birefringent samples. This conclusion is also supported by analyzing the formulation of HPM, which will be discussed next.

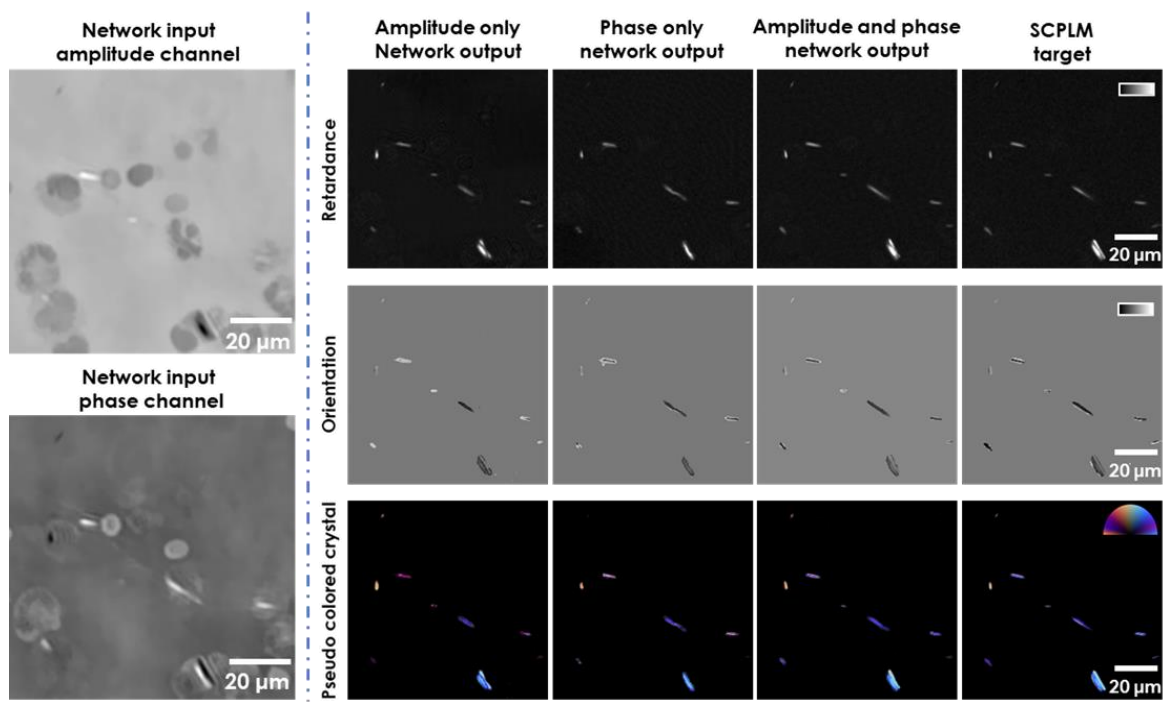

**Figure 6 DL-HPM reconstruction results using different input channels.** The performance of DL-

HPM in general degraded when reducing the input channel to amplitude only or phase only information. Color bar for retardance, from black to white, represents 0 to 1 Rad. Color bar for orientation, from black to white, represents -π/2 to π/2. Color bar for pseudo colored crystals, from left to top to right, represents π, π/2, 0 Rad in the orientation channel. Retardance is represented by the distance from the center of the color bar, ranging from 0 to 0.4.

The evolution of the polarization state in our optical setup (Fig. 1a) can be analyzed using Jones calculus (see Methods section), where for *each pixel* of the reconstructed holographic amplitude and phase images, we can write:

$$\begin{cases} A_{\text{recon}} = \dfrac{\sqrt{a^2 + b^2}}{|\cos\beta - \sin\beta|} \\ \varphi_{\text{recon}} = \text{atan2}(b,a) - \dfrac{\eta}{2} - \dfrac{\pi}{2} - \text{atan2}[0, \cos\beta - \sin\beta] \end{cases} \quad (1)$$

where atan2(*y*, *x*) is the four-quadrant inverse tangent function for point (*x*, *y*), $A_{\text{recon}}$ is the normalized reconstructed amplitude, $\varphi_{\text{recon}}$ is the normalized reconstructed phase (with zero phase in the background), *β* represents the orientation of the linear polarizer with respect to the *x*-axis, and *a* and *b* are defined as:

$$\begin{cases} a = -\cos\beta\sin\eta\sin^2\theta + \sin\beta\sin\eta\cos^2\theta \\ \quad + \cos\beta\cos\theta\sin\theta + \sin\beta\cos\theta\sin\theta \\ \quad - \cos\beta\cos\eta\cos\theta\sin\theta - \sin\beta\cos\eta\cos\theta\sin\theta \\ b = \cos\beta\cos^2\theta - \sin\beta\sin^2\theta \\ \quad + \cos\beta\cos\eta\sin^2\theta - \sin\beta\cos\eta\cos^2\theta \\ \quad - \cos\beta\sin\eta\cos\theta\sin\theta - \sin\beta\sin\eta\cos\theta\sin\theta \end{cases} \quad (2)$$

*θ* represents the orientation of the fast axis of the sample with respect to the *x*-axis, and *η* represents the relative phase retardance.

Unlike the SCPLM method, where the retardance, *η*, and the orientation, *θ*, are encoded in symmetrical equations with a straightforward analytical solution (detailed in the Supporting Information), Equations 1-2 encode the birefringence information in a much more convoluted form. Because of the experimental challenges in obtaining an accurate estimate of *β* as well as the potential phase wrapping related issues, independently solving Equations 1-2 in a pixel-by-pixel manner could result in errors or spatial inconsistencies/artifacts at the output retardance and orientation images. Hence, it elevates the need for a more advanced solution and a robust method such as a deep neural network, which is trained to perform an image-to-image transformation by making use of all the information from multiple pixels within a FOV *simultaneously*. Stated differently, through image data, the deep neural network learns to solve Equations 1-2 over an input FOV, where all the pixels within the complex-valued input image (phase and amplitude) are simultaneously processed to generate the desired output image channels, i.e., the retardance (*η*) and the orientation (*θ*) images.

**Conclusion**

We presented a deep learning-enabled holographic polarization microscope. This framework is advantageous as it only requires the measurement of a single polarization state which can be generated using a simple optical setup, and is capable of accurately reconstructing the quantitative birefringent retardance and orientation information of the specimen. These information channels can dramatically simplify the automatic detection, counting, and classification of birefringent objects within complex media. Powered with these features, our method can be the basis of a rapid point-of-care crystal detection and analysis instrument with automated crystal identification and classification capabilities, which could significantly simplify the clinical procedures used to diagnose diseases related to birefringent crystals, such as gout and pseudogout.

**Methods**

*Lensfree polarization imaging setup*

The presented DL-HPM system utilizes a customized lensfree holographic polarization microscope to capture the input images (Figure 1a). This microscope is able to generate quantitative phase images as well as introduce an intensity contrast to birefringent objects (though it is unable to differentiate high absorbance non-birefringent objects[14]). The microscope uses a laser source filtered by an acousto-optic tunable filter (AOTF) for illumination at 530 nm (~2.5 nm bandwidth). The raw holograms were collected using a CMOS image sensor (IMX 081, Sony, 1.12μm pixel size) at 8 sample to sensor heights, which were used for multi-height phase recovery[32]. A set of low-resolution holograms were captured at 6×6 lateral positions, which were used for pixel-super resolution. Using these images, a high-resolution holographic image was reconstructed and subsequently numerically back propagated[33] to the sample plane using an auto-focusing algorithm[34]. Finally, the reconstructed hologram was normalized to obtain an average background amplitude of 1, and have an average background phase of 0. These normalized images were then passed through the neural network. Details of the holographic image reconstruction techniques including free space propagation, multi-height phase recovery, super resolution, and auto-focusing are presented in the Supporting Information.

To enable detection of the birefringence within the sample, a left-hand polarizer and a customized analyzer were added in the holographic imaging system[14]. Unlike traditional polarization microscopes, where a second circular polarizer (i.e. right-hand polarizer) can be used as the analyzer, holographic imaging systems require background light to form an interference pattern, and the direct use of another circular polarizer would completely reject the background light. Therefore, we designed the analyzer to use a $\lambda/4$ retarder film (75 μm thickness, Edmund Optics), and a linear polarizer (180 μm thickness, Edmund Optics), having the fast axis of the $\lambda/4$ retarder oriented to 25° with respect to the linear polarizer, creating a holographic polarization microscope. These films were affixed directly to the CMOS image sensor using an ultraviolet (UV)-curable adhesive (NOA 68, Norland Products, Cranbury, NJ) as shown in Figure 7b.

*Polarization encoding in the holographic imaging system*

In the analysis of our holographic imaging system, we assume that the sample, polarizer, and analyzer are thin and have negligible gaps between them. We further assume that the light diffracts from the analyzer onto the image sensor after being converted to linearly polarized light by the last layer of the analyzer. Therefore, after its reconstruction, the hologram becomes in-focus at the sample plane (the thicknesses of both the sample and the analyzer are assumed to be negligible).

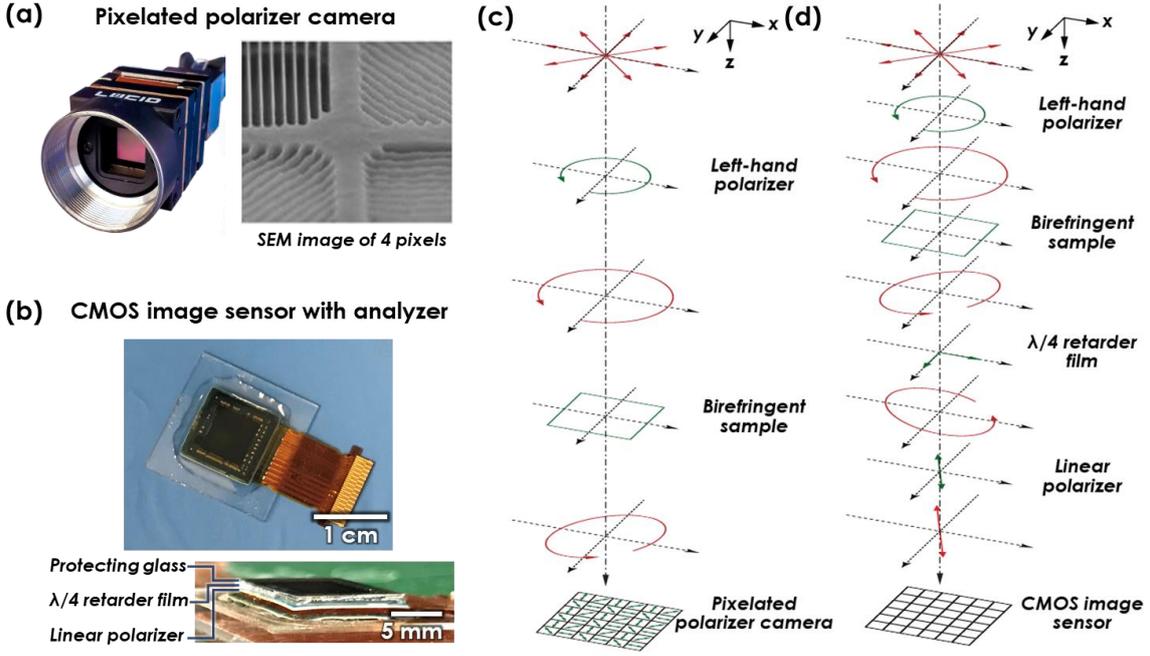

**Figure 7**. **(a) Photo of the four-channel pixelated polarizer camera.** This camera is used in SCPLM. When illuminated with circularly polarized light, four channels with different polarization states are acquired using a single image. **(b) Photo of the CMOS image sensor with a customized analyzer.** This imager is used for DL-HPM. The analyzer film is directly bound to the CMOS image sensor, allowing a certain amount of background light to form the hologram, while also providing one polarization channel to sense the birefringent sample. **(c) Polarization design for SCPLM.** The red plot represents the light polarization state at each plane. The green plot represents an optical component that changes the polarization state. **(d) Polarization design for DL-HPM.**

The evolution of the polarization in our imaging system can be analyzed using Jones calculus[35]. The light field in the presented framework was designed to be transmitted through a left-hand polarizer, birefringent sample, $\lambda/4$ retardation plate and a linear polarizer (Fig. 2d). Each of these optical components can be formulated as:

I. Input left-hand circularly polarized (LHCP) light:

$$\mathbf{E}_{in} = \frac{1}{\sqrt{2}} \begin{bmatrix} 1 \\ -i \end{bmatrix} \quad (3)$$

where $i^2 = -1$, and LHCP is defined *from the point of view of the source*.

II. Birefringent sample:

$$\mathbf{M}_{sample} = e^{-\frac{i\eta}{2}} \begin{bmatrix} \cos^2\theta + e^{i\eta}\sin^2\theta & (1-e^{i\eta})\cos\theta\sin\theta \\ (1-e^{i\eta})\cos\theta\sin\theta & \sin^2\theta + e^{i\eta}\cos^2\theta \end{bmatrix} \quad (4)$$

where $\theta$ represents the orientation of the fast axis of the sample with respect to the *x*-axis, and $\eta$ represents the relative phase retardance.

III. $\lambda/4$ retarder:

$$\mathbf{M}_{\text{retarder}} = e^{-\frac{i\pi}{4}} \begin{bmatrix} \cos^2\alpha + i\sin^2\alpha & (1-i)\cos\alpha\sin\alpha \\ (1-i)\cos\alpha\sin\alpha & \sin^2\alpha + i\cos^2\alpha \end{bmatrix} \quad (5)$$

where $\alpha$ represents the orientation of the fast axis of the $\lambda/4$ retarder with respect to the *x*-axis.

IV. Linear polarizer:

$$\mathbf{M}_{\text{linear}} = \begin{bmatrix} \cos^2\beta & \cos\beta\sin\beta \\ \cos\beta\sin\beta & \sin^2\beta \end{bmatrix} \quad (6)$$

where $\beta$ represents the orientation of the linear polarizer with respect to the *x*-axis.

The output light field can then be expressed as:

$$\mathbf{E}_{\text{out}} = \mathbf{M}_{\text{linear}} \mathbf{M}_{\text{retarder}} \mathbf{M}_{\text{sample}} \mathbf{E}_{\text{in}} \quad (7)$$

Equations 1-2 reported in the Discussion section are obtained by rearranging Equation 7, and applying a background normalization step (detailed in the Supporting Information).

*Dataset preparation*

The neural networks were trained using image pairs captured using both SCLPM and the holographic imaging systems. We used 6 clinical MSU samples for training, and 2 additional MSU samples for testing. To ensure that the network training can generalize to new samples, the slides used to train the neural network were chosen to have different concentrations of MSU crystals (example FOVs of each slide are shown in Supporting Information Figure S1). In addition, a single TCA sample was used for both training and testing (where blind testing was performed on new regions); similarly, a single starch sample was used for blind testing.

In order to train the neural network to learn the image transformation from a lensfree holographic imaging modality to a lens-based SCPLM system, an accurately co-registered training dataset is required. This co-registration begins by bicubic down-sampling the target polarization images by a factor of 0.345/0.373 (obtained with the benchtop microscope) to match the pixel size of the holographic microscope; the ground truth images were created using SCPLM with an effective pixel size of 0.345 μm, while the holographic images used as the network input have an effective pixel size of 0.373 μm. Next, a rough matching between the two sets of images is obtained by finding the overlapping area with maximum correlation between the amplitude channels of the images. Once the images have been roughly aligned, global matching is performed by applying an affine transformation calculated using MATLAB's multimodal image registration framework[36]. This framework extracts features from the amplitude channel of the images and matches them with the affine matrix. Next, large fields of view were cropped, and matched to each other using an elastic pyramidal registration algorithm, which allows for pixel level matching[37] based upon the amplitude channel of the images. As discussed above, these holograms were normalized to have an average background amplitude of 1, and an average background phase of 0, before being passed through the neural network. This normalization step helps the neural network to be applied more consistently to different samples. The orientation of areas of the polarization images without any birefringence are set to zero to eliminate noise in the labels. This is done by setting the orientation value of any pixel (below a threshold) in the corresponding retardance channel to zero.

*Neural network*

A Generative Adversarial Network (GAN) framework was used to perform the image transformation reported in this paper. In addition to this GAN loss, a mean absolute error ($L_1$) loss was used to ensure that the transformation is accurate, and a total variation (TV) loss is used as a regularization term. GANs use two separate networks for their operation. A generator network (G(·)) is used to transform the holographic images into their polarization counterparts. A second network known as the discriminator (D(·)) is used to discriminate between the ground truth images ($z$) and the generated images (G($x$)). The two networks learn from one another, with the generator gradually learning how to create images that match the feature distribution of the target dataset, while the discriminator gets better at distinguishing between ground truth images and their generated counterparts.

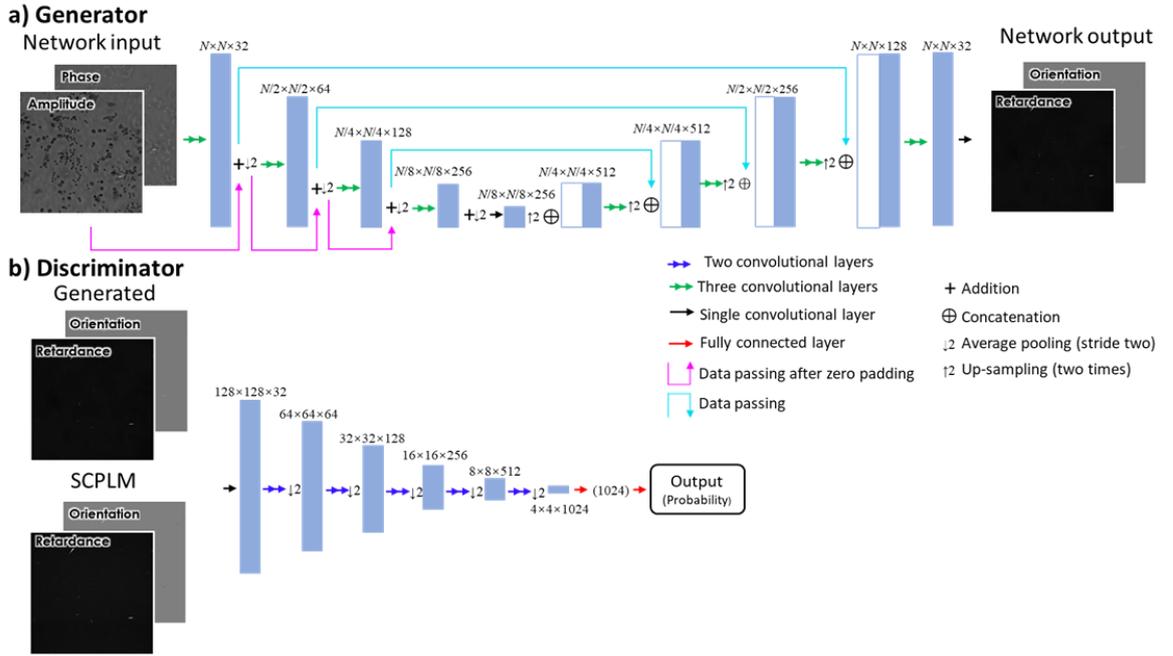

**Figure. 8 Network architecture.** (a) Diagram of the generator portion of the network. (b) Diagram of the discriminator portion of the network.

The overall loss function can be described as:

$$l_{generator} = L_1\{G(x),z\} + \lambda_1 \times TV\{G(x)\} + \lambda_2 \times (1-D(G(x)))^2 \quad (8)$$

where $\lambda_1$ and $\lambda_2$ are constants used to balance the various terms of the loss function. The $L_1$ loss was balanced to make up ~25% of the total loss function, while the total variation loss makes up ~0.5% of the overall loss function. The $L_1$ loss is defined as:

$$L_1\{G(x),z\} = \frac{1}{M \times N \times K} \sum_k \sum_i \sum_j |z_{i,j,k} - G(x)_{i,j,k}| \quad (9)$$

where $K$ is the number of image channels ($k$ =1 represents the retardance channel, and $k$ =2 represents the orientation channel), $M$ and $N$ are the number of pixels on each axis, and $i$ and $j$ represent the pixel indices of the image. The total variation loss is defined as:

$$\text{TV}(G(x)) = \frac{1}{M \times N \times K} \sum_k \sum_i \sum_j |G(x)_{i+1,j,k} - G(x)_{i,j,k}| + |G(x)_{i,j+1,k} - G(x)_{i,j,k}| \qquad (10)$$

In order to train the discriminator a separate loss function is used, defined as:

$$l_{\text{discriminator}} = D(G(x))^2 + (1 - D(z))^2 \qquad (11)$$

The generator network uses the U-net architecture[38], as shown in Figure 8(a). This U-net begins with a convolutional layer increasing the number of channels to 32, and is made up of four "down-blocks" followed by four "up-blocks". Each down-block consists of three convolutional layers, which together double the number of channels. These layers are followed by an average pooling layer with a kernel size and stride of 2. After these down-blocks, an equal number of up-blocks are applied. The up-blocks begin by bilinear up-sampling the images and similarly apply three convolutional layers, and reduce the number of channels by a factor of four. Between the two sets of blocks, skip connections are added. The skip connections allow small scale data to pass through the network, avoiding the effects of the down-sampling by concatenating the output of each down-block with the input to each up-block. Following these blocks, a convolutional layer reduces the number of channels to two, which match the two channels of the polarization images.

The discriminator network (Figure 8(b)) receives the generated images or the SCPLM images, and attempts to distinguish between the two. The discriminator is first made up of a convolutional layer which increases the number of channels from 3 to 32. This is followed by five blocks, each made up of two convolutional layers, the second of which doubling the number of channels and using a stride of two. Following these five blocks are two fully connected layers, which reduce the image to a single number which can have a sigmoid function applied to it.

Each convolutional layer uses a kernel size of 3×3 and is followed by the leaky rectified linear unit (LeakyReLU) activation function which is defined as:

$$\text{LeakyReLU} = \begin{cases} x & \text{for } x > 0 \\ 0.1x & \text{otherwise} \end{cases} \qquad (12)$$

In the training phase, we used the adaptable movement estimation (Adam) optimizer, with a learning rate of $10^{-4}$ for the generator, and $10^{-5}$ for the discriminator. The network begins with the generator being trained 7 times for each training of the discriminator, with this ratio being reduced by 1 every 4000 iterations down to a minimum of 3. The network was trained for 30,000 iterations of the discriminator. Training was performed using a single 2080 Ti (Nvidia), with Python version 3.6.0 and TensorFlow Version 1.11.0. MATLAB version R2018a was used for preprocessing.

**Acknowledgements.** The authors acknowledge the support of National Institutes of Health (NIH, R21AR072946).

**References**

(1) Wolman, M. Polarized Light Microscopy as a Tool of Diagnostic Pathology. *Journal of Histochemistry & Cytochemistry* **1975**, *23* (1), 21–50.

(2) Arun Gopinathan, P.; Kokila, G.; Jyothi, M.; Ananjan, C.; Pradeep, L.; Humaira Nazir, S. Study of Collagen Birefringence in Different Grades of Oral Squamous Cell Carcinoma Using Picrosirius Red and Polarized Light Microscopy. *Scientifica* **2015**, *2015*.

(3) Vijaya, B.; Dalal, B. S.; Manjunath, G. V. Primary Cutaneous Amyloidosis: A Clinico-Pathological Study with Emphasis on Polarized Microscopy. *Indian Journal of Pathology and Microbiology* **2012**, *55* (2), 170.

(4) Jin, L.-W.; Claborn, K. A.; Kurimoto, M.; Geday, M. A.; Maezawa, I.; Sohraby, F.; Estrada, M.; Kaminksy, W.; Kahr, B. Imaging Linear Birefringence and Dichroism in Cerebral Amyloid Pathologies. *Proceedings of the National Academy of Sciences* **2003**, *100* (26), 15294–15298.

(5) Cornwell III, G. G.; Murdoch, W. L.; Kyle, R. A.; Westermark, P.; Pitkänen, P. Frequency and Distribution of Senile Cardiovascular Amyloid: A Clinicopathologic Correlation. *The American journal of medicine* **1983**, *75* (4), 618–623.

(6) Mccarty, D. J.; Hollander, J. L. Identification of Urate Crystals in Gouty Synovial Fluid. *Ann. Intern. Med.* **1961**, *54*, 452–460. https://doi.org/10.7326/0003-4819-54-3-452.

(7) McGill, N. W.; Dieppe, P. A. Evidence for a Promoter of Urate Crystal Formation in Gouty Synovial Fluid. *Ann. Rheum. Dis.* **1991**, *50* (8), 558–561. https://doi.org/10.1136/ard.50.8.558.

(8) Gatter, R. A. Editorial: The Compensated Polarized Light Microscope in Clinical Rheumatology. *Arthritis Rheum.* **1974**, *17* (3), 253–255. https://doi.org/10.1002/art.1780170308.

(9) Gordon, C.; Swan, A.; Dieppe, P. Detection of Crystals in Synovial Fluids by Light Microscopy: Sensitivity and Reliability. *Ann. Rheum. Dis.* **1989**, *48* (9), 737–742. https://doi.org/10.1136/ard.48.9.737.

(10) Park, J. W.; Ko, D. J.; Yoo, J. J.; Chang, S. H.; Cho, H. J.; Kang, E. H.; Park, J. K.; Song, Y. W.; Lee, Y. J. Clinical Factors and Treatment Outcomes Associated with Failure in the Detection of Urate Crystal in Patients with Acute Gouty Arthritis. *Korean J. Intern. Med.* **2014**, *29* (3), 361–369. https://doi.org/10.3904/kjim.2014.29.3.361.

(11) Oldenbourg, R.; Mei, G. New Polarized Light Microscope with Precision Universal Compensator. *Journal of microscopy* **1995**, *180* (2), 140–147.

(12) Oldenbourg, R. Polarization Microscopy with the LC-PolScope. *Live Cell Imaging: A Laboratory Manual* **2005**, 205–237.

(13) Kim, Y.; Jeong, J.; Jang, J.; Kim, M. W.; Park, Y. Polarization Holographic Microscopy for Extracting Spatio-Temporally Resolved Jones Matrix. *Optics Express* **2012**, *20* (9), 9948–9955.

(14) Zhang, Y.; Lee, S. Y. C.; Zhang, Y.; Furst, D.; Fitzgerald, J.; Ozcan, A. Wide-Field Imaging of Birefringent Synovial Fluid Crystals Using Lens-Free Polarized Microscopy for Gout Diagnosis. *Scientific reports* **2016**, *6*, 28793.

(15) Bai, B.; Wang, H.; Liu, T.; Rivenson, Y.; FitzGerald, J.; Ozcan, A. Pathological Crystal Imaging with Single-Shot Computational Polarized Light Microscopy. *Journal of biophotonics* **2020**, *13* (1), e201960036.

(16) Oh, C.; Isikman, S. O.; Khademhosseinieh, B.; Ozcan, A. On-Chip Differential Interference Contrast Microscopy Using Lensless Digital Holography. *Opt. Express, OE* **2010**, *18* (5), 4717–4726. https://doi.org/10.1364/OE.18.004717.

(17) Greenbaum, A.; Luo, W.; Su, T.-W.; Göröcs, Z.; Xue, L.; Isikman, S. O.; Coskun, A. F.;

Mudanyali, O.; Ozcan, A. Imaging without Lenses: Achievements and Remaining Challenges of Wide-Field on-Chip Microscopy. *Nature Methods* **2012**, *9* (9), 889–895. https://doi.org/10.1038/nmeth.2114.

(18) Bishara, W.; Su, T.-W.; Coskun, A. F.; Ozcan, A. Lensfree On-Chip Microscopy over a Wide Field-of-View Using Pixel Super-Resolution. *Opt. Express, OE* **2010**, *18* (11), 11181–11191. https://doi.org/10.1364/OE.18.011181.

(19) Mudanyali, O.; Tseng, D.; Oh, C.; Isikman, S. O.; Sencan, I.; Bishara, W.; Oztoprak, C.; Seo, S.; Khademhosseini, B.; Ozcan, A. Compact, Light-Weight and Cost-Effective Microscope Based on Lensless Incoherent Holography for Telemedicine Applications. *Lab Chip* **2010**, *10* (11), 1417–1428. https://doi.org/10.1039/c000453g.

(20) Ozcan, A.; McLeod, E. Lensless Imaging and Sensing. *Annual Review of Biomedical Engineering* **2016**, *18* (1), 77–102. https://doi.org/10.1146/annurev-bioeng-092515-010849.

(21) Zhu, H.; Isikman, S. O.; Mudanyali, O.; Greenbaum, A.; Ozcan, A. Optical Imaging Techniques for Point-of-Care Diagnostics. *Lab Chip* **2012**, *13* (1), 51–67. https://doi.org/10.1039/C2LC40864C.

(22) de Haan, K.; Rivenson, Y.; Wu, Y.; Ozcan, A. Deep-Learning-Based Image Reconstruction and Enhancement in Optical Microscopy. *Proceedings of the IEEE* **2020**, *108* (1), 30–50. https://doi.org/10.1109/JPROC.2019.2949575.

(23) Barbastathis, G.; Ozcan, A.; Situ, G. On the Use of Deep Learning for Computational Imaging. *Optica, OPTICA* **2019**, *6* (8), 921–943. https://doi.org/10.1364/OPTICA.6.000921.

(24) Sinha, A.; Lee, J.; Li, S.; Barbastathis, G. Lensless Computational Imaging through Deep Learning. *Optica, OPTICA* **2017**, *4* (9), 1117–1125. https://doi.org/10.1364/OPTICA.4.001117.

(25) Rivenson, Y.; Zhang, Y.; Günaydın, H.; Teng, D.; Ozcan, A. Phase Recovery and Holographic Image Reconstruction Using Deep Learning in Neural Networks. *Light: Science & Applications* **2018**, *7* (2), 17141. https://doi.org/10.1038/lsa.2017.141.

(26) Wu, Y.; Rivenson, Y.; Zhang, Y.; Wei, Z.; Günaydin, H.; Lin, X.; Ozcan, A. Extended Depth-of-Field in Holographic Imaging Using Deep-Learning-Based Autofocusing and Phase Recovery. *Optica* **2018**, *5* (6), 704–710.

(27) Liu, T.; Wei, Z.; Rivenson, Y.; de Haan, K.; Zhang, Y.; Wu, Y.; Ozcan, A. Deep Learning-Based Color Holographic Microscopy. *Journal of biophotonics* **2019**, e201900107.

(28) Liu, T.; de Haan, K.; Rivenson, Y.; Wei, Z.; Zeng, X.; Zhang, Y.; Ozcan, A. Deep Learning-Based Super-Resolution in Coherent Imaging Systems. *Scientific reports* **2019**, *9* (1), 3926.

(29) Rivenson, Y.; Liu, T.; Wei, Z.; Zhang, Y.; de Haan, K.; Ozcan, A. PhaseStain: The Digital Staining of Label-Free Quantitative Phase Microscopy Images Using Deep Learning. *Light: Science & Applications* **2019**, *8* (1), 23. https://doi.org/10.1038/s41377-019-0129-y.

(30) Christiansen, E. M.; Yang, S. J.; Ando, D. M.; Javaherian, A.; Skibinski, G.; Lipnick, S.; Mount, E.; O'Neil, A.; Shah, K.; Lee, A. K.; Goyal, P.; Fedus, W.; Poplin, R.; Esteva, A.; Berndl, M.; Rubin, L. L.; Nelson, P.; Finkbeiner, S. In Silico Labeling: Predicting Fluorescent Labels in Unlabeled Images. *Cell* **2018**, *173* (3), 792-803.e19. https://doi.org/10.1016/j.cell.2018.03.040.

(31) Wu, Y.; Luo, Y.; Chaudhari, G.; Rivenson, Y.; Calis, A.; de Haan, K.; Ozcan, A. Bright-Field Holography: Cross-Modality Deep Learning Enables Snapshot 3D Imaging with Bright-Field Contrast Using a Single Hologram. *Light: Science & Applications* **2019**, *8* (1), 25.

(32) Greenbaum, A.; Ozcan, A. Maskless Imaging of Dense Samples Using Pixel Super-Resolution Based Multi-Height Lensfree on-Chip Microscopy. *Opt. Express, OE* **2012**, *20* (3), 3129–3143.

(33) Goodman, J. W. *Introduction to Fourier Optics*; Roberts and Company Publishers, 2005.

(34) Zhang, Y.; Wang, H.; Wu, Y.; Tamamitsu, M.; Ozcan, A. Edge Sparsity Criterion for Robust Holographic Autofocusing. *Opt Lett* **2017**, *42* (19), 3824–3827.

(35) Jones, R. C. A New Calculus for the Treatment of Optical SystemsI. Description and Discussion of the Calculus. *J. Opt. Soc. Am., JOSA* **1941**, *31* (7), 488–493. https://doi.org/10.1364/JOSA.31.000488.

(36) Register Multimodal MRI Images - MATLAB & Simulink Example https://www.mathworks.com/help/images/registering-multimodal-mri-images.html (accessed May 29, 2020).

(37) Rivenson, Y.; Ceylan Koydemir, H.; Wang, H.; Wei, Z.; Ren, Z.; Günaydın, H.; Zhang, Y.; Gorocs, Z.; Liang, K.; Tseng, D. Deep Learning Enhanced Mobile-Phone Microscopy. *Acs Photonics* **2018**, *5* (6), 2354–2364.

(38) Ronneberger, O.; Fischer, P.; Brox, T. U-Net: Convolutional Networks for Biomedical Image Segmentation. *arXiv:1505.04597 [cs]* **2015**.